# Comments on «Second caloric virial coefficients for real gases and combined spherical symmetric potential for simple molecular interactions»

I.H. Umirzakov

*Institute of Thermophysics, acad. Lavrenteva Ave., 1, Novosibirsk, 630090, Russia*

*e-mail: cluster125@gmail.com*

## Abstract

It is shown that the interaction potentials for argon, krypton, xenon, carbon dioxide, ammonia, water, n-pentane, n-octane, 1-propanol, suggested in [1] cannot describe experimental second density virial coefficient data within their uncertainty over a wide temperature range.

1. Comment on statement of [1] that «the attraction between polar molecules at large distances might be expressed as varying with $U(r) = -cr^{-6} - \mu T^{-1} r^{-6}$, which describes the induction and spontaneous part of dipole-dipole interactions. For fluids with zero dipole moment such as argon, carbon dioxide, and *n*-alkanes $U(r) = -cr^{-6}$».

The potential $U(r) = -cr^{-6} - \mu T^{-1} r^{-6}$ must describe **dispersion forces of London**, the induction and spontaneous part of dipole-dipole interactions [2]. So $c = c_6 + 2\alpha\mu_d^2$, $\mu = 2\mu_d^4 / 3k$, where $c_6$ is the parameter of dispersion forces of London, $\mu_d$ - the electrical dipole moment of the molecule and $\alpha$ is the polarizability of the molecule, $2\alpha\mu_d^2$ corresponds to the induction. For fluids with zero dipole moment such as argon, carbon dioxide, and *n*-alkanes $U(r) = -c_6 r^{-6}$.

2. It is necessary to check the correctness of using the potential $U(r,T) = -\mu T^{-1} r^{-6}$ for calculating the second density virial coefficient by Eq. (5) of [1]. The potential $U(r,T)$ is obtained by integration over all orientations of the polar molecules interacting via anisotropic potential [2].

The second density virial coefficient calculated from (5) for the anisotropic potential of two identical point dipoles placed in the center of hard spheres with diameter D is equal to [2]

$$B(T) = \frac{2\pi}{3} ND^3 [1 - \frac{1}{3} (\mu_d^2 / D^3 kT)^2 - \sum_{m=2}^{\infty} G_m \{(\mu_d^4 / D^6 (kT)^2\}^m / (2m)!(2m-1)] .$$

The second density virial coefficient calculated from (5) of [1] for the potential $U(r,T) = -\mu T^{-1} r^{-6}$ , where $\mu = 2\mu_d^4 / 3k$, is equal to

$$B(T) = \frac{2\pi}{3} ND^3 [1 - \frac{2}{3}(\mu_d^2 / D^3 kT)^2 - \sum_{m=2}^{\infty} \{2\mu_d^4 / 3D^6 (kT)^2\}^m / m!(2m-1)].$$

One can see that $G_m \neq (2/3)^m (2m)!/m!$, so the second density virial coefficients calculated above are not identical. Therefore the potential $V(r,T) = -\mu T^{-1} r^{-6}$ cannot be used to calculate second virial coefficient.

It is easy to see that the potential $V(r,T) = -\mu T^{-1} r^{-6}$ cannot be used to calculate other virial coefficients.

3. As one can see from table 2 of [1] the parameters $\sigma$ and $d$ of the potentials of substances considered obey $\sigma \approx d$. The parameter $\sigma$ is approximately equal to the size of the molecule. But the potentials $U(r) = -cr^{-6} - \mu T^{-1} r^{-6}$ and $U(r) = -c_6 r^{-6}$ are valid for $r >> \sigma$.

4. Comments on the statement of [1] that «for fluids with a strong spontaneous dipole moment, such as water, $cr^{-6} << \mu T^{-1} r^{-6}$».

One can see from the Table 120 [2] that $cr^{-6} \approx \mu T^{-1} r^{-6}$ for temperatures T=800-1000K. The temperatures up to 1000K are considered for water (see Fig. 2 (c), Fig. 3 (b) and Fig. 5 (a) in [1]).

5. The formulae (31) –(37) of [1] have many mistakes, and they must be replaced by

$$A = -c_2 T_B \{\exp(-\beta / kT_B) - 1\} - c_1 T_B \{\exp(\varepsilon / kT_B) - 1\}. \quad (31)$$

$$B(T) = (T_B / T)[c_2 \{\exp(-\beta / kT_B) - 1\} + c_1 \{\exp(\varepsilon / kT_B) - 1\}] - \\ -[c_2 \{\exp(-\beta / kT) - 1\} + c_1 \{\exp(\varepsilon / kT) - 1\}], \quad (32)$$

$$B(T) = (T_B / T)[c_2 \{\exp(-\beta / kT_B) - 1\} + c_1 \{\exp(\varepsilon / kT_B) - 1\}] - \\ -[c_2 \{\exp(-\beta / kT) - 1\} + c_1 \{\exp(\varepsilon / kT) - 1\}] - D(T_B - T) / T_B T^2, \quad (33)$$

$$B(T) = (T_B^2 / T^2)[c_2 \{\exp(-\beta / kT_B) - 1\} + c_1 \{\exp(\varepsilon / kT_B) - 1\}] - \\ -[c_2 \{\exp(-\beta / kT) - 1\} + c_1 \{\exp(\varepsilon / kT) - 1\}], \quad (34)$$

$$B(T) = (T_B / T)[c_2 \{\exp(-\beta / kT_B) - 1\} + c_1 \{\exp(T_C / T_B) - 1\}] - \\ -[c_2 \{\exp(-\beta / kT) - 1\} + c_1 \{\exp(T_c / T) - 1\}], \quad (35)$$

$$B(T) = (T_B / T)[c_2 \{\exp(-\beta / kT_B) - 1\} + c_1 \{\exp(T_C / T_B) - 1\}] - \\ -[c_2 \{\exp(-\beta / kT) - 1\} + c_1 \{\exp(T_C / T) - 1\}] - D(T_B - T) / T_B T^2, \quad (36)$$

$$B(T) = (T_B^2 / T^2)[c_2 \{\exp(-\beta / kT_B) - 1\} + c_1 \{\exp(T_C / T_B) - 1\}] - \\ -[c_2 \{\exp(-\beta / kT) - 1\} + c_1 \{\exp(T_C / T) - 1\}], \quad (37)$$

6. The sense of the parameter $\gamma$ in table 2 in [1] is not known. We establish the sense of this parameter: $\gamma = c/d^6$. It is also necessary to note that $c = 2\mu_d^4/3kd^6$ in the table 2 for polar molecules: water, ammonia and 1-propanol. The senses of the parameters $c$ (for polar molecules) and $\gamma$ are not established in [1].

7. Comments on the statements of [1] that:

- «The parameters of the interaction potential …. were determined from experimental second density virial coefficients. This potential describes second density virial coefficients *B(T)* within there experimental uncertainties over a wide temperature range»;

- «The model described above for the intermolecular potential function for polar fluids contains … . To apply the present potential function model for real fluids … . A nonlinear optimization method was used which minimized the statistic $\chi^2 = [\sum_{i=1}^{N}\{B_i(\exp) - B_i(calc)\}^2/\sigma_i^2]/(N-m)$, where ….. $B_i(calc)$ is the corresponding second density virial coefficient value calculated from the model. The results obtained are given in table 2. In figures …. we compare density and heat capacity second virial coefficients for …. calculated from the present model with recommended values from the literature and present experimental results. The comparison shows remarkably good agreement between calculated and experimental values of second density virial coefficients. The root mean square errors are: for argon, 0.9 $cm^3 \cdot mol^{-1}$; for water, 3.4 $cm^3 \cdot mol^{-1}$; for carbon dioxide, 8.7 $cm^3 \cdot mol^{-1}$; for *n*-pentane, 26.2 $cm^3 \cdot mol^{-1}$; for *n*-octane, 24.8 $cm^3 \cdot mol^{-1}$; for 1-propanole, 6.2 $cm^3 \cdot mol^{-1}$. Therefore, this potential function allows experimental second density virial coefficient data to be described within their uncertainty for a wide class of fluids with various molecular structures over a wide temperature range»;

- «The new improved combined intermolecular potential model is able to represent the experimental density and heat capacity second virial coefficients almost to within their experimental uncertainties. This model of the potential function is useful for interpolating and extrapolating the experimental results».

We have shown that the statements do not correspond to the reality (for all substances from table 2 [1] except neon). It one can see from results presented below.

We have calculated the second density virial coefficient using Eq. (5), the model potential functions (28) and (29) with parameters from table 2 of [1]. We put *c*=0 in (29) as assumed in [1]. The results are presented in figures 1-10, where the temperature (T2 or Ta in K) dependence of the relative deviations of value of second density virial coefficient (SDVC) Btheor calculated from experimental (tabulated) data (B2 or Ba) of [3-5]. The root mean square errors are: for neon 0.628 (1) $cm^3 \cdot mol^{-1}$; for krypton 26.314 (10) $cm^3 \cdot mol^{-1}$; for xenon 6.666 (5) $cm^3 \cdot mol^{-1}$; for argon 8.516 (5) $cm^3 \cdot mol^{-1}$; for water 96.108 (16.89) $cm^3 \cdot mol^{-1}$; for carbon dioxide 160.06 (6.6) $cm^3 \cdot mol^{-1}$; for *n*-pentane 523.619 (60) $cm^3 \cdot mol^{-1}$; for *n*-octane 1209 (140) $cm^3 \cdot mol^{-1}$; for 1-propanole 465.203 (50) $cm^3 \cdot mol^{-1}$, for ammonia 73.399 (5.109) $cm^3 \cdot mol^{-1}$, where the maximal values of the uncertainties of experimental (tabulated) data are shown in the brackets.

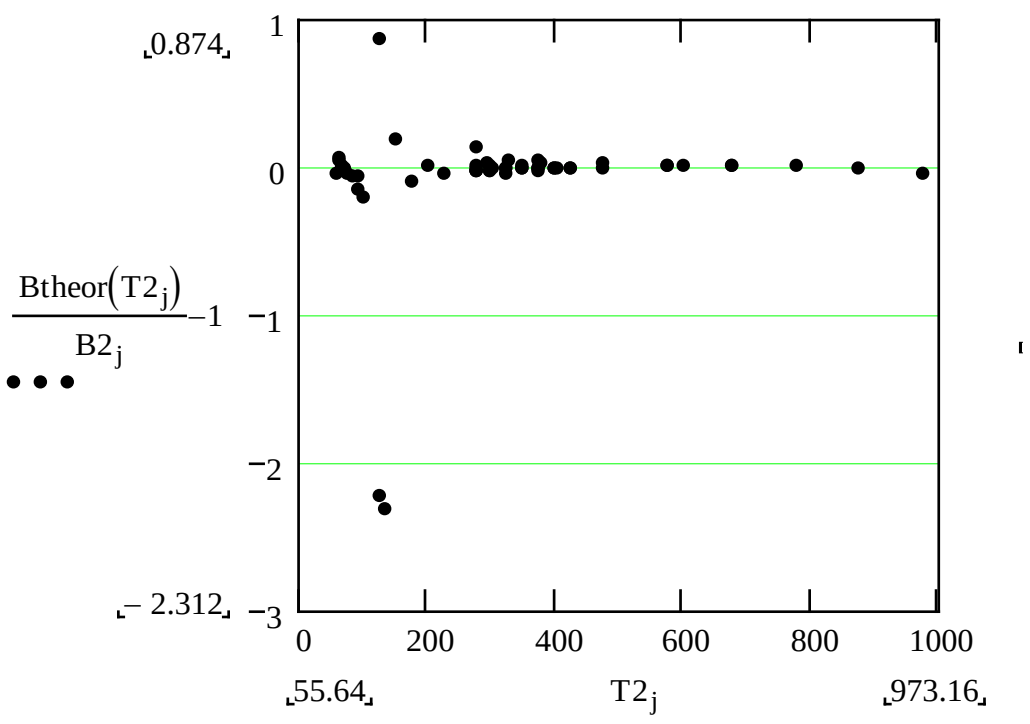

Fig. 1a. Neon. Temperature dependence of the relative deviations of SDVC Btheor calculated from data of [3].

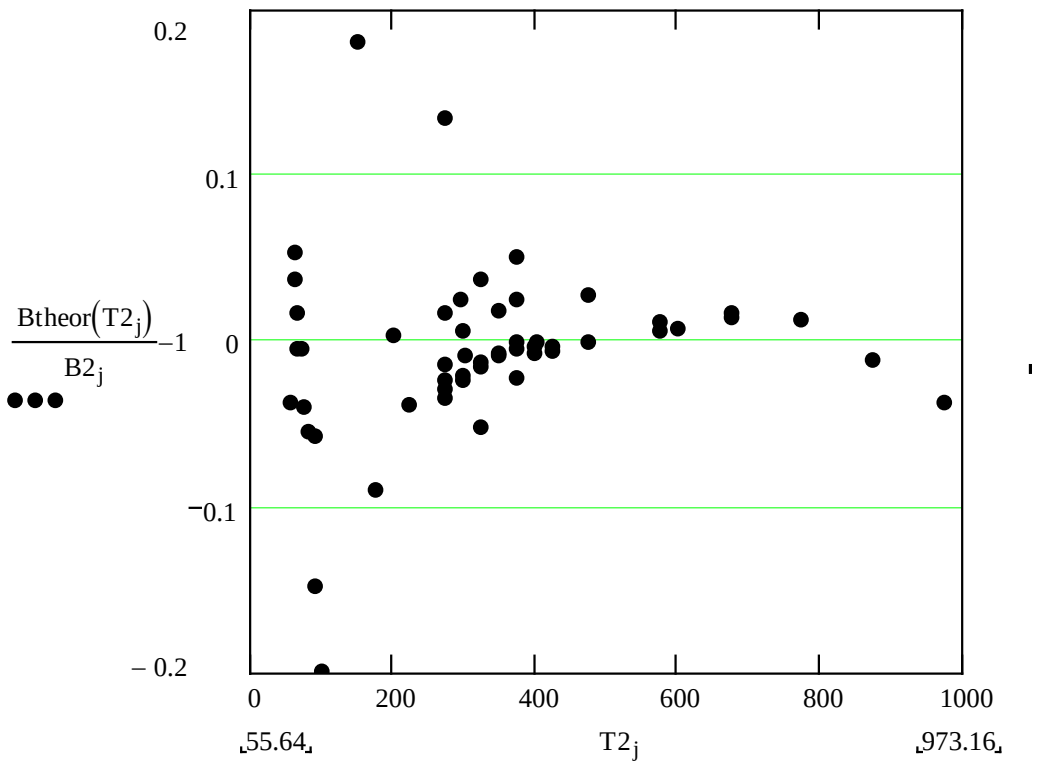

Fig. 1b. Neon Temperature dependence of the relative deviations of SDVC Btheor calculated from data of [3].

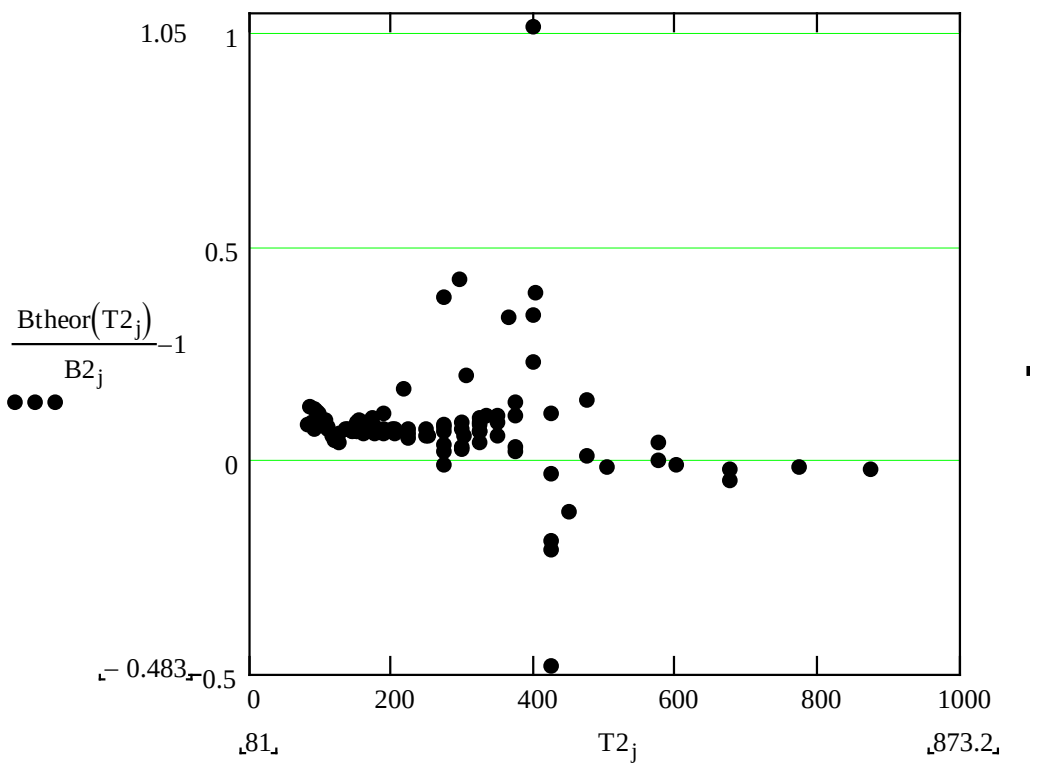

Fig. 2. Argon. Temperature dependence of the relative deviations of SDVC Btheor calculated from data of [3]

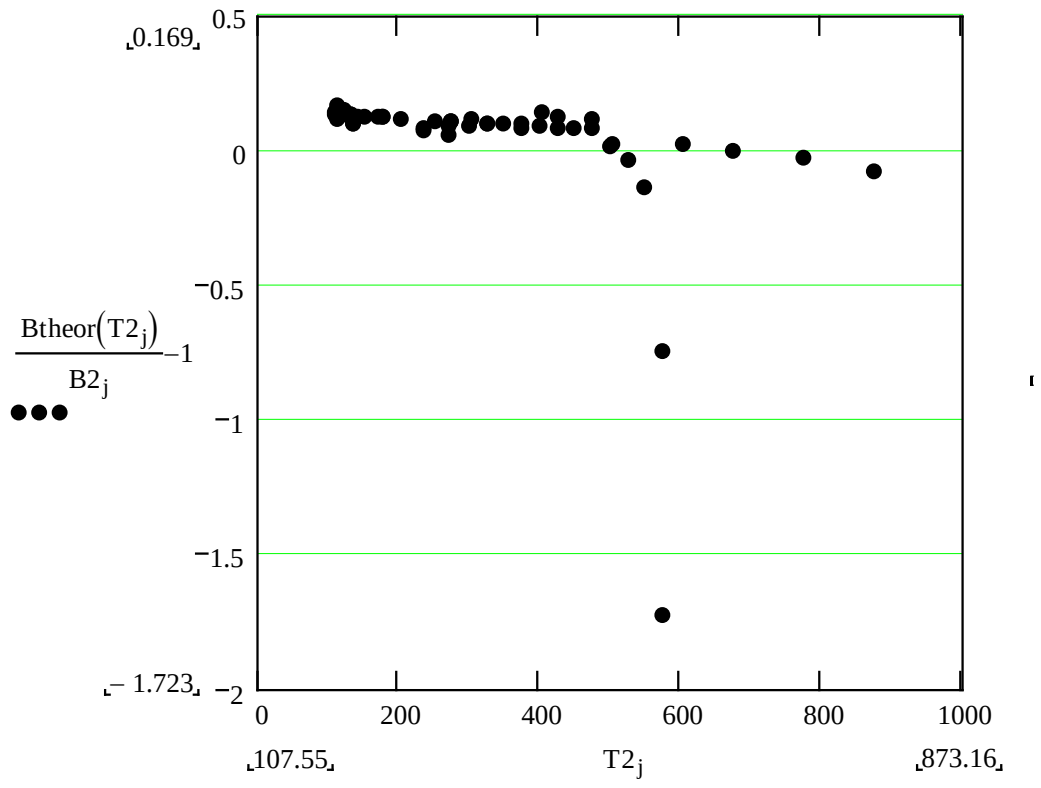

Fig. 3a. Kripton. Temperature dependence of the relative deviations of SDVC Btheor calculated from data of [3].

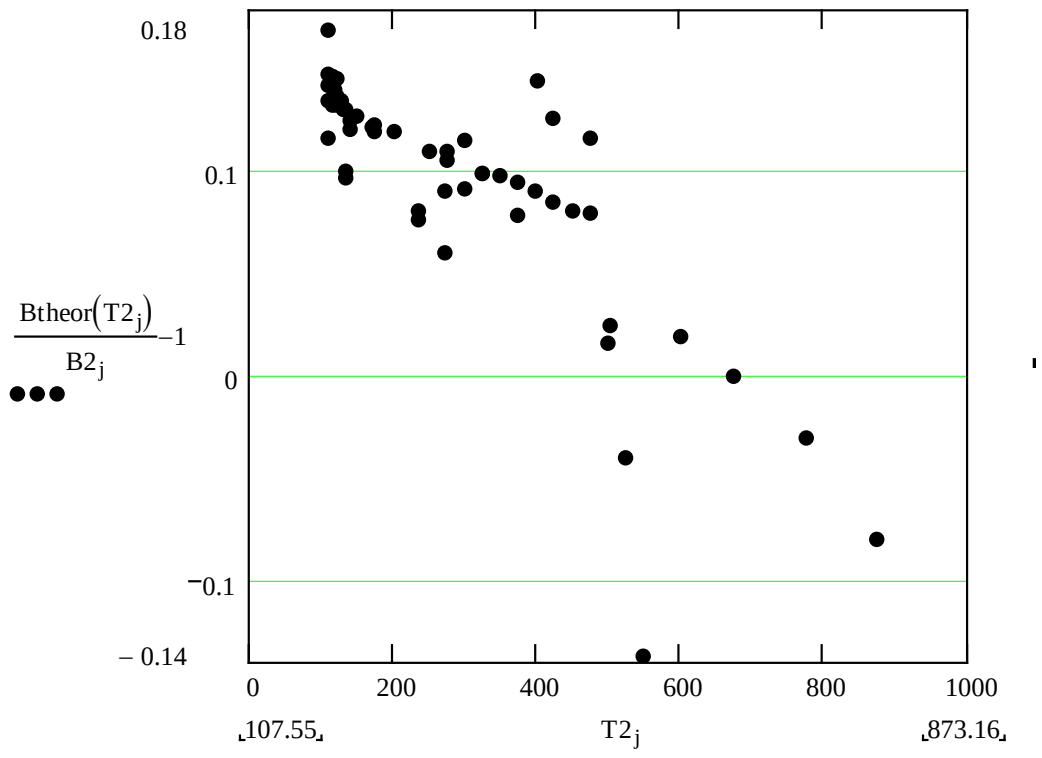

Fig. 3b. Kripton. Temperature dependence of the relative deviations of SDVC Btheor calculated from data of [3].

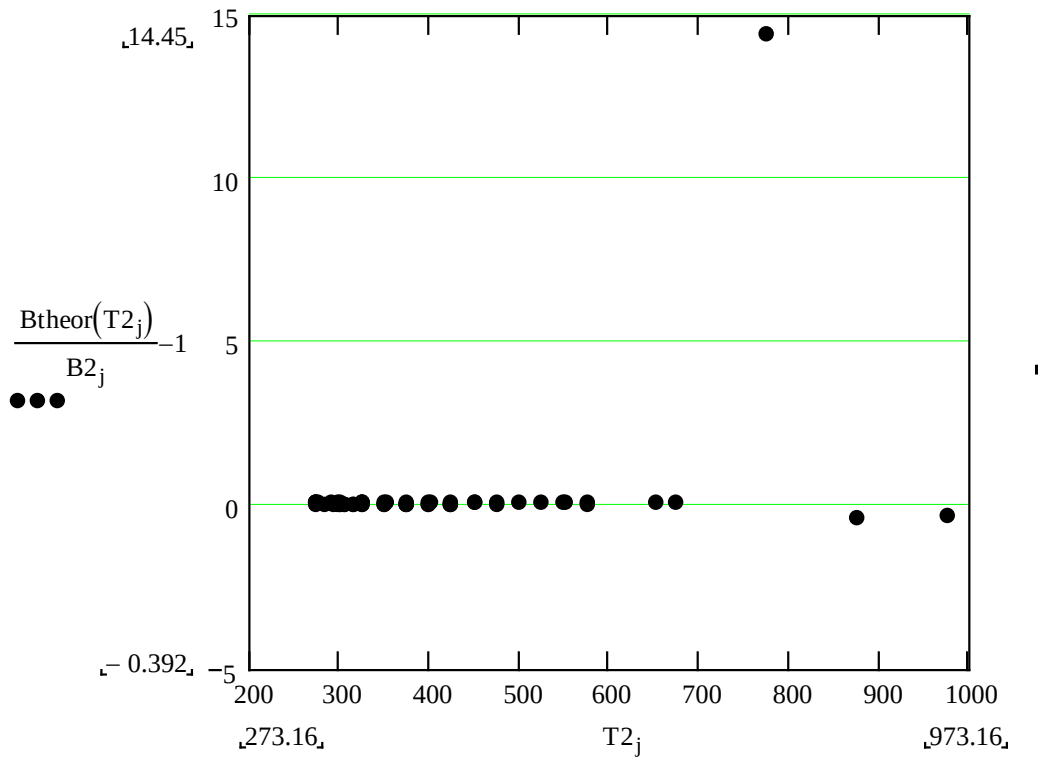

Fig. 4a. Xenon. Temperature dependence of the relative deviations of SDVC Btheor calculated from data of [3]

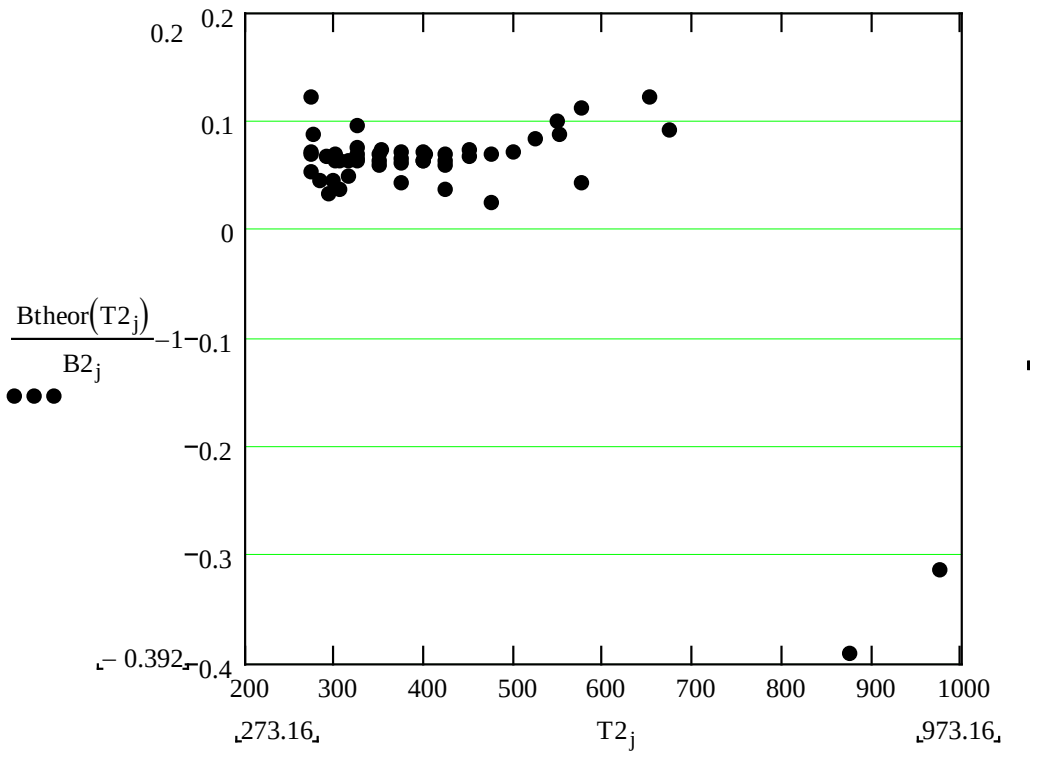

Fig. 4b. Xenon. Temperature dependence of the relative deviations of SDVC Btheor calculated from data of [3].

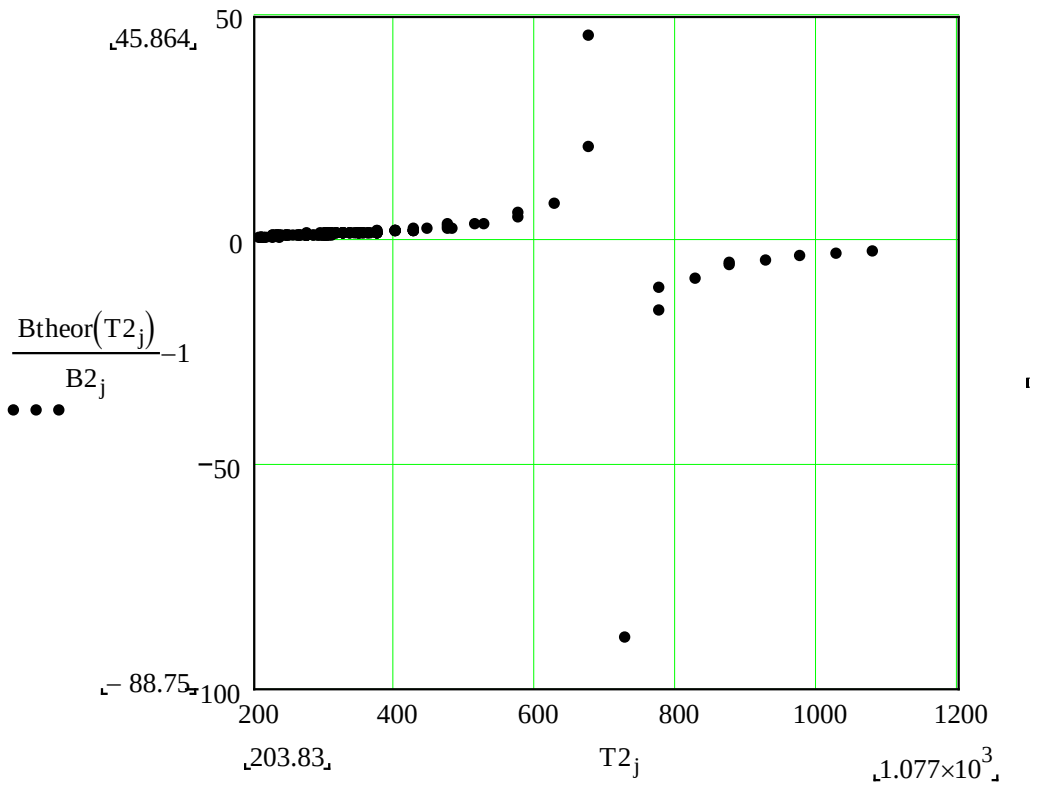

Fig. 5a. Carbon dioxide. Temperature dependence of the relative deviations of SDVC Btheor calculated from data of [3].

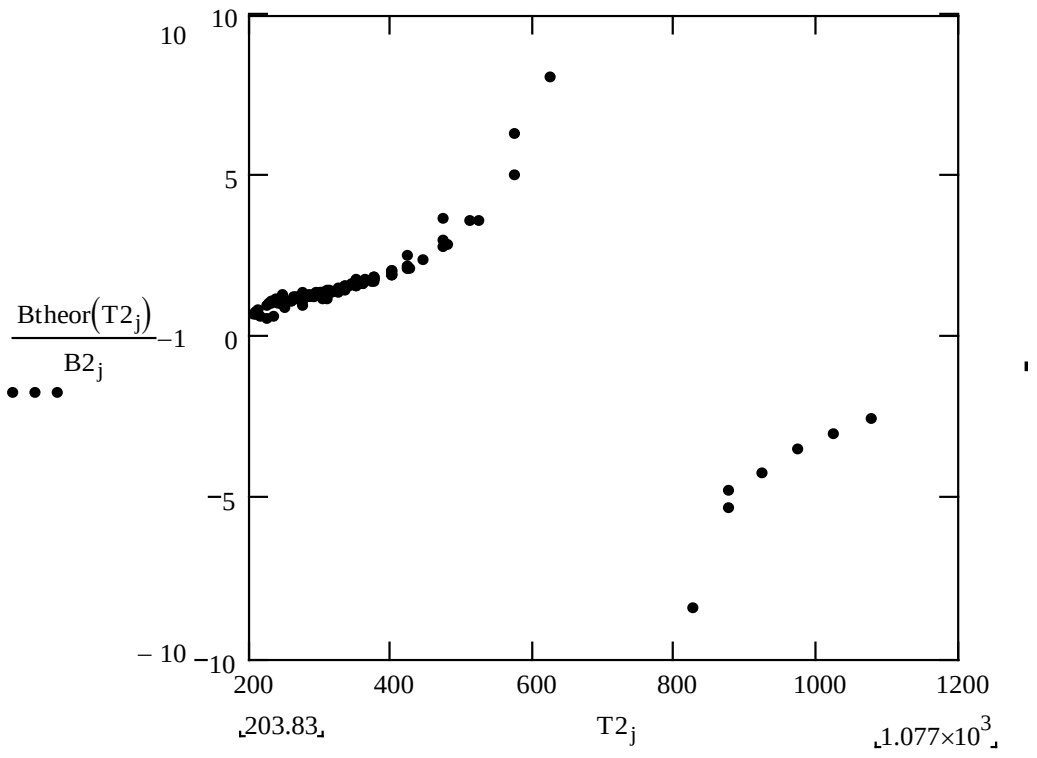

Fig. 5b. Carbon dioxide. Temperature dependence of the relative deviations of SDVC Btheor calculated from data of [3].

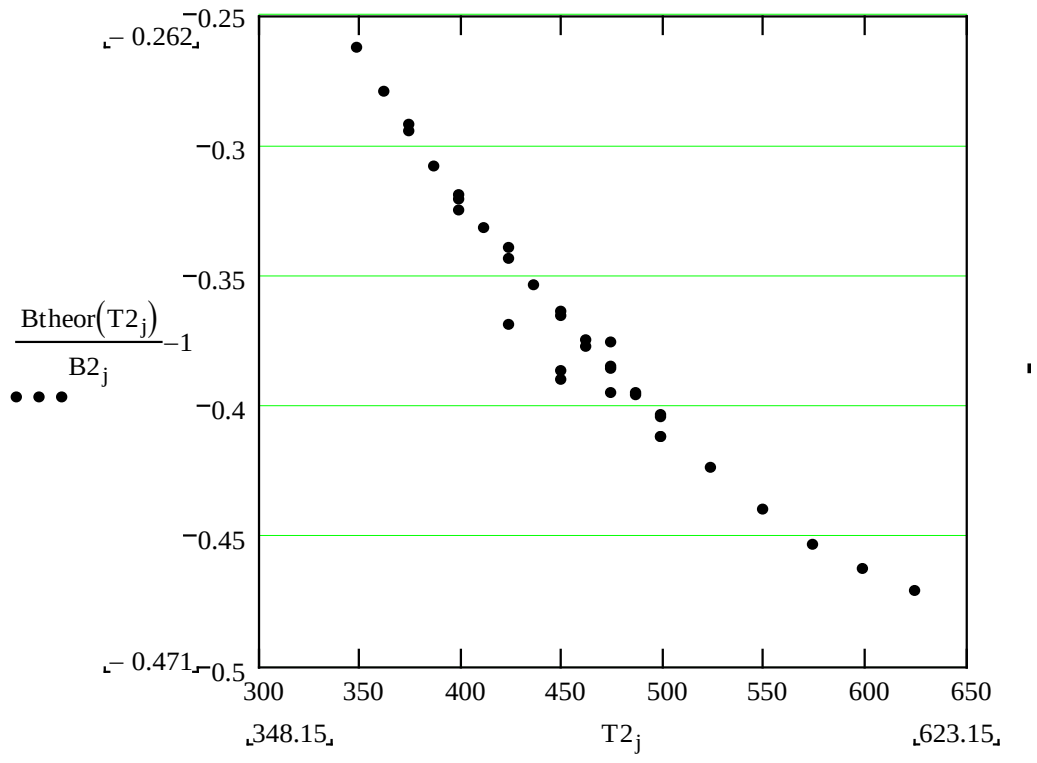

Fig. 6. Water. Temperature dependence of the relative deviations of SDVC Btheor calculated from data of [3].

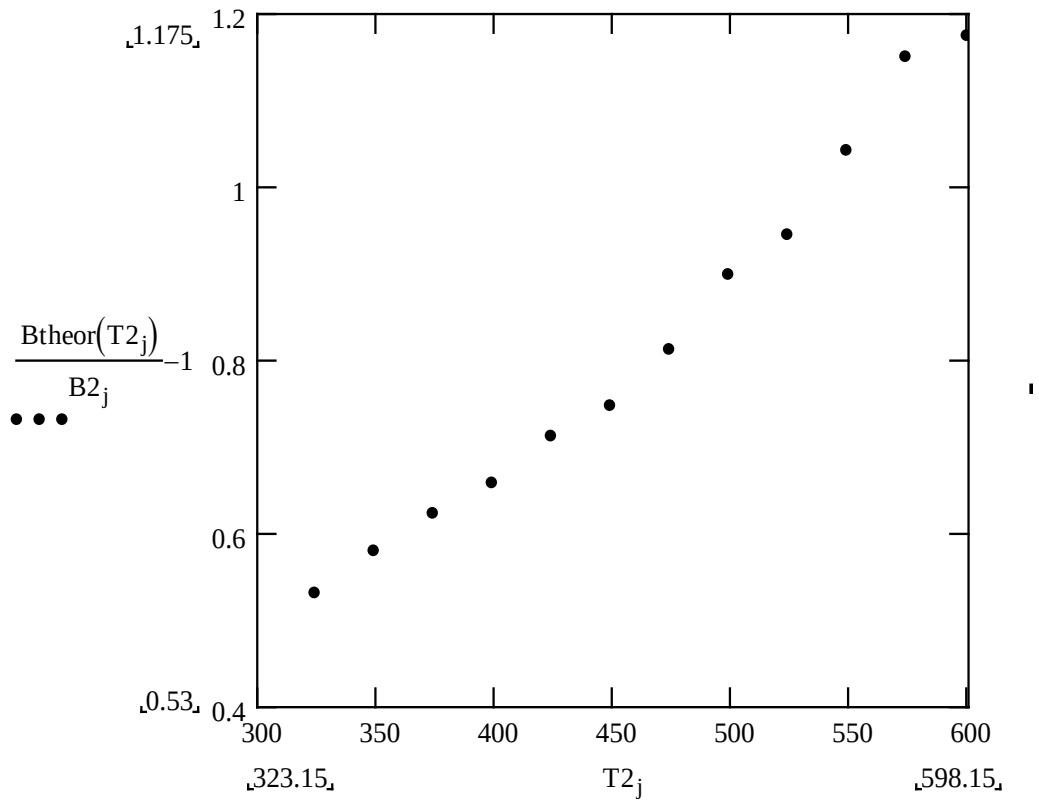

Fig. 7. Ammonia. Temperature dependence of the relative deviations of SDVC Btheor calculated from data of [3].

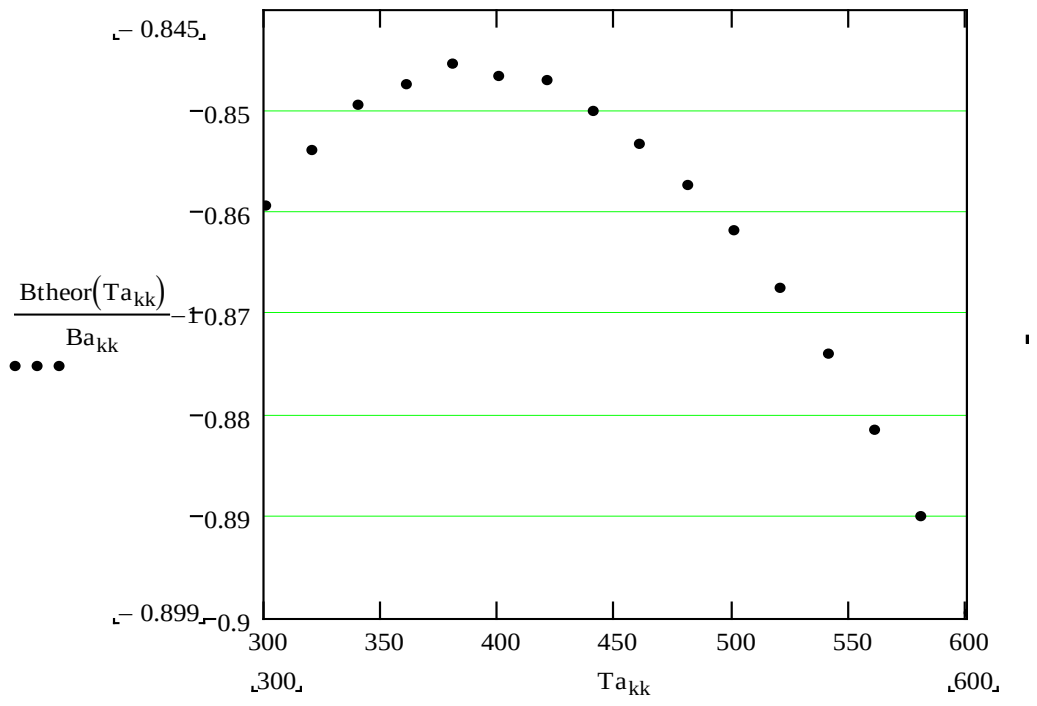

Fig. 8. n-pentane. Temperature dependence of the relative deviations of SDVC Btheor calculated from data of [4].

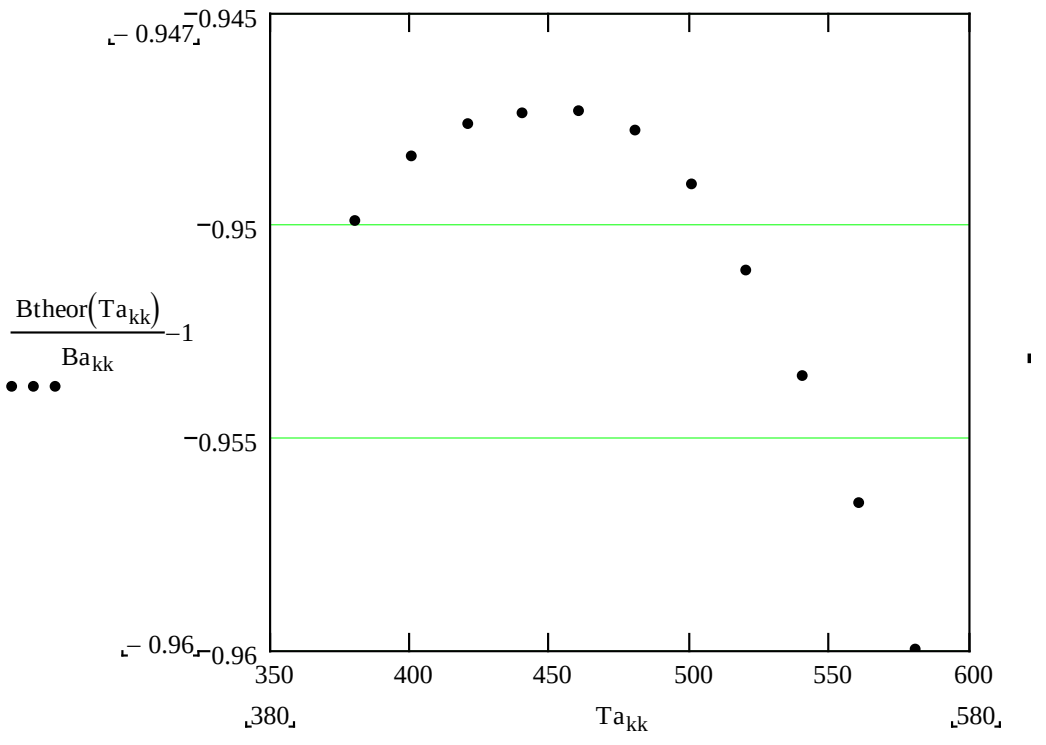

Fig.9. n-octane. Temperature dependence of the relative deviations of SDVC Btheor calculated from data of [4].

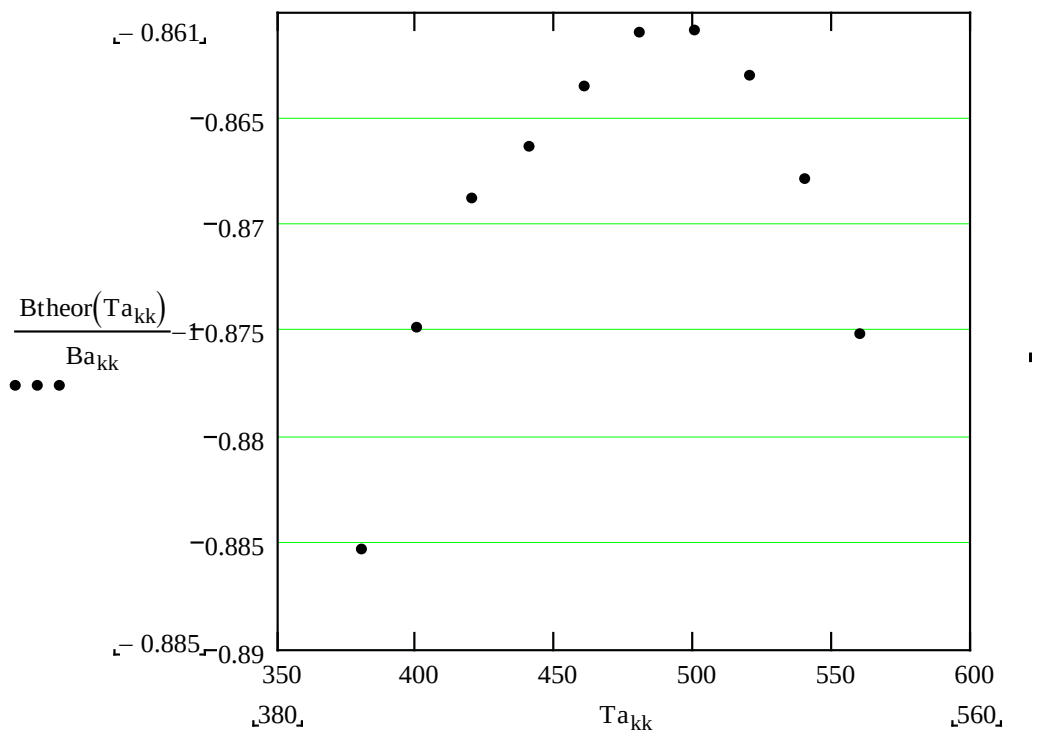

Fig. 10. 1-propanol. Temperature dependence of the relative deviations of SDVC Btheor calculated from data of [5].